\documentstyle[12pt]{article}
\input{epsf}

\newcommand{\bea}{\begin{eqnarray}}
\newcommand{\eea}{\end{eqnarray}}
\newcommand{\beq}{\begin{equation}}
\newcommand{\eeq}{\end{equation}}
\newcommand{\del}{\partial}
\newcommand{\lishi}{\langle\!\langle}
\newcommand{\rishi}{\rangle\!\rangle}

\begin{document}
\thispagestyle{empty}




\renewcommand{\thefootnote}{\fnsymbol{footnote}}
\title{Logarithmic conformal field theory with boundary\footnotemark[2]}

\author{Shinsuke Kawai\\\\
Theoretical Physics, Department of Physics, University of Oxford,\\
1 Keble Road, Oxford OX1 3NP, United Kingdom\\
{\tt kawai@thphys.ox.ac.uk}}
\date{\today}
\maketitle

\begin{abstract}
This lecture note covers topics on boundary conformal field theory, 
modular transformations and the Verlinde formula, and boundary logarithmic CFT.
An introductory review on CFT with boundary and a discussion of its 
applications to logarithmic cases are given. 
LCFT at $c=-2$ is mainly discussed. 
\end{abstract}

\noindent
\footnotetext[2]{This article is based on the lectures at the International 
Summer School on Logarithmic Conformal Field Theory and Its Applications, 
Sept. 2001, IPM, Tehran, I.R. Iran.}
\newpage
\tableofcontents
\newpage


\section{Introduction}

Conformal field theories with logarithmic correlation functions have been
studied actively for the past several years.
Such theories arise naturally as generalisations of the well-investigated
unitary Virasoro minimal theories or WZNW theories with integral level,
and are believed to have many applications in statistical models and
string / brane physics.
These logarithmic conformal field theories (LCFTs) were investigated 
sporadically by several authors\cite{knizhnik,rozansky,saleur,frenkel} in the 
late eighties and early nineties, and systematic study started with Gurarie's 
work\cite{gurarie} in 1993.  
By now various models, e.g. $c=-2$ model\cite{saleur,gurarie,rational,local}, 
gravitationally dressed CFTs\cite{bilalkogan1,bilalkogan2},
WZNW models with fractional $k$\cite{frenkel,fractional} 
and $k=0$\cite{caux1,caux2,nichols1,nichols2} have been studied, and a number 
of applications, including critical polymers\cite{saleur,dupsal,ivashke}, 
percolation\cite{carperc,watts}, quantum Hall effect\cite{graflonay,ino},
disordered systems\cite{caux1,caux2,caux3,maassarani,guruswamy}, 
sandpile model\cite{mahieu}, turbulence\cite{turb1,turb2,turb3}, MHD\cite{mhd},
D-brane recoil\cite{dbrecoil1,dbrecoil2}, etc. have been discussed.
Readers are referred to the other lecture notes\cite{rtabar,mflohr,mgaberdiel} 
of the summer school for more complete historical account and description on
the state of the art of the study of LCFT. 

This article is intended to give an overview of basic concepts in boundary
conformal field theory, and to present recent attempts to apply them 
to logarithmic theories. 
Motivation for considering boundaries in CFTs is somewhat obvious when we try 
to model statistical systems:
any existing sample of material has finite extent and a theory on the 
infinite plane is only approximately valid.
In order to model the system near the boundary where the finite-size effect 
is not negligible, we need to define CFT on a topology with boundary. 
Boundary CFT is also essential for string theory. In the past decade
higher-dimensional objects called branes have attracted much
attention. 
Branes are defined as end points of open strings, and are described using 
boundary CFTs. 
The study of branes is developing rapidly, and after the emergence of the 
brane-world scenario, phenomenological interest is growing as well.
Apart from these `physical' aspects, boundary CFT is an attractive subject 
because of its beautiful mathematical structure.
It is well known that the modular invariance of partition functions
on the torus leads to the classification of rational conformal theories.
For the conformal theories with boundary, the modular invariance gives a
classification of boundaries which may be realised in a physical system. 
These modular properties arise from the fact that characters of rational
CFTs happen to be linear representation of the modular group. 

Boundary logarithmic conformal theory is admittedly rather a young subject,
and the known results so far depend on a few simple models.
This article presents mainly two topics: the first is on properties of
boundary correlation functions\cite{koganwheater}, and the second is on 
the classification of boundary states based on modular 
invariance\cite{kawaiwheater}. 
We shall discuss these for the $c=-2$ model, after reviewing basic ideas of
boundary CFT applied to non-logarithmic examples.
These results are based on straightforward generalisation of standard concepts 
in boundary CFT to
the simplest logarithmic theory, and more discussions such as those based on
some concrete physical models are desirable.
Nevertheless such an attempt is clearly a natural first step to the thorough 
understanding of logarithmic theories, and we believe these topics are 
important for the development of non-unitary conformal theories in general.

This article is organised as follows. In the next section we review
some basic concepts and techniques in ordinary boundary conformal field theory.
In particular, derivation of boundary correlation functions (Subsec.2.2) and
Cardy's classification of boundary states (Subsec.2.3) are given in detail.
We also mention Verlinde formula and boundary operators.
In Sec. 3 we apply these ideas to $c=-2$ logarithmic theory with boundary,
where boundary correlation functions and Cardy's classification are discussed
mainly.
We summarise, conclude, and give some prospects in Sec. 4. 
As the discussion of Sec.3.4 is similar to more familiar Ising model ($c=1/2$)
case, we summarise the Majorana fermion construction of the 
boundary states of the Ising model in App.A.
Some formulae on elliptic functions are collected in App.B. 
\newpage


\section{Conformal field theory with boundary}


In this section we review standard techniques and concepts in non-logarithmic
boundary conformal field theory.
What we have in mind is simple diagonal unitary minimal models, 
such as the Ising model.
We start, in the first subsection, by discussing conformal invariance in the 
presence of a boundary.
In Subsec.2.2 we review the mirroring method\cite{cardy1} for finding boundary 
correlation functions.
We describe in Subsec.2.3 the classification of consistent boundary states 
based on the modular invariance, which is known as Cardy's fusion 
method\cite{cardy2}.
The relation between the modular transformation matrix and the fusion rule 
(Verlinde formula\cite{verlinde}) and its relevance in the boundary 
theory\cite{cardy2} is reviewed in Subsec.2.4.
In Subsec.2.5 we discuss boundary operators\cite{cardy2,cardylewellen} and 
give an example of a statistical model\cite{carperc} where the concept of 
boundary operators plays a central role.


\subsection{Conformal transformation with boundary}

Let us start by considering what is meant by conformal invariance
in the presence of a boundary. Let
\beq
ds^2=g_{\mu\nu}(x)dx^\mu dx^\nu,
\eeq
be the line element of the manifold we work on.
Since the metric is a tensor, it transforms as
\beq
g^{\mu\nu}(x)\rightarrow\tilde g^{\mu\nu}(\tilde x)
=\frac{\del\tilde x^\mu}{\del x^\lambda}
\frac{\del\tilde x^\nu}{\del x^\rho}
g^{\lambda\rho}(x).
\eeq
The conformal transformation is defined as a mapping which preserves
the metric $g^{\mu\nu}(x)$ up to a scale factor,
\beq
g^{\mu\nu}(x)\rightarrow\tilde g^{\mu\nu}(\tilde x)
=\Omega(x)^2g^{\mu\nu}(x).
\eeq
In two dimensions this condition is written as
\beq
\left(\frac{\del\tilde x^0}{\del x^0}\right)^2
+\left(\frac{\del\tilde x^0}{\del x^1}\right)^2
=\left(\frac{\del\tilde x^1}{\del x^0}\right)^2
+\left(\frac{\del\tilde x^1}{\del x^1}\right)^2,
\eeq
\beq
\frac{\del\tilde x^0}{\del x^0}\frac{\del\tilde x^1}{\del x^0}
+\frac{\del\tilde x^0}{\del x^1}\frac{\del\tilde x^1}{\del x^1}=0,
\eeq
which are equivalent either to 
\beq
\frac{\del \tilde x^0}{\del x^0}=\frac{\del\tilde x^1}{\del x^1},\;\;
\frac{\del \tilde x^1}{\del x^0}=-\frac{\del\tilde x^0}{\del x^1},
\label{eqn:creq}
\eeq
or to
\beq
\frac{\del \tilde x^0}{\del x^0}=-\frac{\del\tilde x^1}{\del x^1},\;\;
\frac{\del \tilde x^1}{\del x^0}=\frac{\del\tilde x^0}{\del x^1}.
\label{eqn:anticreq}
\eeq
These are the Cauchy-Riemann equations and their antiholomorphic counterpart.
Defining $z=x^0+ix^1$ and $\bar z=x^0-i x^1$,
we conclude that the conformal transformation in two-dimensions (without 
considering boundary) is equivalent to
analytic mapping on the complex plane\cite{difrancesco}.  

On the full plane, the conformal mapping
\bea
&&z\rightarrow w(z)=\sum_n a_n z^n,\\
&&\bar z\rightarrow \bar w(\bar z)=\sum_n \bar a_n \bar z^n,
\eea
is generated by an infinite number of generators $a_n$ and $\bar a_n$,
which imposes strict constraints on the field theory.
In a geometry with boundary, we may take the line $x^1=0$ as the boundary and 
consider a CFT on the upper half plane. 
As the field theory is restricted to a fixed geometry, the conformal
transformation must keep the boundary $x^1=0$ invariant.
This means
\beq
\mbox{Im}\;w(x)|_{x^1=0}=0\;\;\Leftrightarrow\;\;w(x^0)=\bar w(x^0)
\;\;\Leftrightarrow\;\;a_n=\bar a_n.
\eeq
Although the number of generators is reduced by half due to this condition,
we still have an infinite dimensional conformal group and conformal
invariance remains extremely powerful\cite{aos}.  
Note that the holomorphic and antiholomorphic generators are coupled on the 
boundary. 
This allows us to interpret the antiholomorphic part as an analytic 
continuation of the holomorphic part, as we shall see in the next subsection.


\subsection{Boundary correlation functions}

The existence of null vectors in minimal CFTs allows us to find $n$-point 
correlation functions as solutions to differential 
equations of hypergeometric type\cite{bpz}.
This method was generalised to CFTs on the half plane by Cardy\cite{cardy1},
using the mirroring technique which is familiar in electrostatics.
In this subsection we review this method and find the spin correlation
functions of the Ising model on the upper half plane.

The behaviour of correlation functions under the conformal transformations 
is described by the conformal Ward identities.
For a CFT on the upper half plane they are
\beq
\delta\langle\phi_1\phi_2\cdots\rangle\ =
\frac{-1}{2\pi i}\oint_{C}dz\epsilon\langle 
T(z)\phi_1\phi_2\cdots\rangle
+\frac{1}{2\pi i}\oint_{C} d\bar z\bar\epsilon\langle 
\bar T(\bar z)\phi_1\phi_2\cdots\rangle,
\label{eqn:wardid1}
\eeq
as
$z\rightarrow w=z+\epsilon$, $\bar z\rightarrow \bar w =\bar z+\bar\epsilon$,
and
$\bar\epsilon = \epsilon^*$.
The contours are the semicircle $C$ which encircles all the coordinates 
$(z_i,\bar z_i)$ of the operators (Fig.1a).
Since there is no energy-momentum flow across the boundary, the 
energy-momentum tensor satisfies the condition
\beq
\left[T-\bar T\right]_{z=\bar z}=0,
\label{eqn:t=tbar}
\eeq
on the boundary $z=\bar z$. 
This condition also means the diffeomorphism invariance of the boundary
as the conformal transformation is generated by the energy-momentum tensor.
We can use the condition (\ref{eqn:t=tbar}) to extend the domain of definition
of $T(z)$, by mapping the antiholomorphic part on the upper half plane (UHP) 
to the holomorphic part on the lower half plane (LHP), as
$T(z^*)=\bar T(\bar z)$.
The antiholomorphic dependence of the correlation function on the UHP 
coordinates is similarly mapped to the holomorphic dependence on the
LHP coordinates.
The antiholomorphic part of the Ward identities (\ref{eqn:wardid1}) is then 
mapped into the holomorphic part on the LHP, as shown in Fig.1b.
The direction of the integration contour on the LHP is reversed (Fig.1c) by 
changing the sign of the second term in (\ref{eqn:wardid1}). 
Since the two contours along the boundary cancel each other, the contours can 
be
concatenated to make contour of full circle (Fig.1d), leading to a
much simpler conformal Ward identity,
\beq
\delta\langle\phi_1\phi_2\cdots\rangle
=\frac{-1}{2\pi i}\oint_{C-C^*} dz\epsilon(z)\langle T(z)
\phi_1(z_1)\bar\phi_1(z_1^*)\phi_2(z_2)\bar\phi_2(z_2^*)\cdots\rangle.
\eeq
This means that the $n$-point function on the UHP satisfy the same differential
equation as the {\em chiral} $2n$-point function on the full plane, with
the LHP coordinates obtained through mirroring with respect to the boundary.

\begin{figure}
~
\epsfxsize=45mm
\epsfysize=70mm
\epsffile{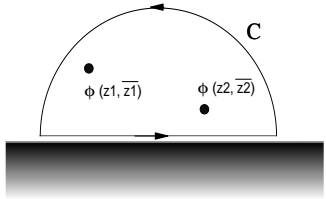}
~~~~~~~~~~~~~~~~~~~~~~~~~~~~
\epsfxsize=45mm
\epsfysize=70mm
\epsffile{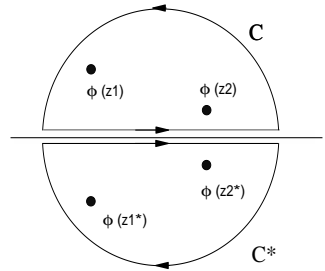}\\
(a) Contour $C$ on UHP. ~~~~~~~~~~~~~~~~~~~~~~~~~~~~~~
(b) Mirroring: $C\rightarrow C^*$.\\
\epsfxsize=45mm
\epsfysize=70mm
\epsffile{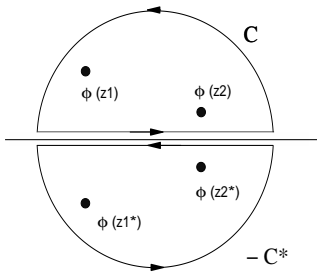}
~~~~~~~~~~~~~~~~~~~~~~~~~~~~
\epsfxsize=45mm
\epsfysize=70mm
\epsffile{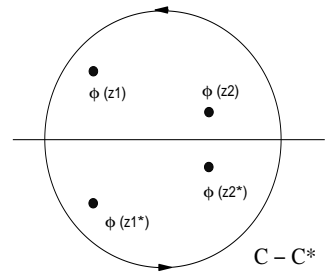}\\
(c) Reverse the direction. ~~~~~~~~~~~~~~~~~~~~~~~~~~~
(d) Merge two contours.
\caption{The antiholomorphic coordinate dependence of CFT on the UHP (a) is 
mapped to the holomorphic dependence on the LHP by mirroring (b). 
Flipping the direction of the contour on the LHP (c), and merging the two 
contours, the Ward identity of the $n$-point function on the UHP is shown
to be equivalent to that of the $2n$-point function on the full plane.}
\end{figure}

Let us see this in the example of the Ising model, and find the spin-spin
correlation function on the UHP.
As the boundary $2$-point function on the half plane is equivalent to the 
$4$-point function on the full plane, one may write
\beq
\langle\sigma(z_1,\bar z_1)\sigma(z_2,\bar z_2)\rangle_{UHP}
=\langle\sigma(z_1)\sigma(z_2)\sigma(z_1^*)\sigma(z_2^*)
\rangle_{chiral},
\eeq
where $\sigma$ is the spin operator.
Due to the existence of a singular vector at level 2, the $4$-point function of
$\sigma=\phi_{1,2}$ satisfies a second order differential equation,
\beq
\left\{\partial_z^2-\frac 34\sum_{i=1}^3
\left[\frac{1}{z-z_i}\partial_{z_i}+\frac{1/16}{(z-z_i)^2}\right]
\right\}\langle\sigma(z)\sigma(z_1)\sigma(z_2)\sigma(z_3)\rangle=0.
\eeq
Using the global conformal transformations, this partial differential equation
reduces to a hypergeometric differential equation which is solved as
\beq
\langle\sigma(z_1,\bar z_1)\sigma(z_2,\bar z_2)\rangle_{UHP}
=\left[
\frac{(z_1-z_1^*)(z_2-z_2^*)}{|z_1-z_2|^2|z_1-z_2^*|^2}\right]^{1/8}
\left\{A\sqrt{\sqrt{1-\eta}+1}+B\sqrt{\sqrt{1-\eta}-1}\right\},
\eeq
where
$\eta=(z_1-z_2)(z_1^*-z_2^*)/(z_1-z_1^*)(z_2-z_2^*)
=-|z_1-z_2|^2/4\mbox{Im} z_1 \mbox{Im} z_2$ is the cross ratio,
which takes a negative real value $-\infty<\eta<0$ in the physical region.

The coefficients $A$ and $B$ are to be determined by the boundary conditions.
It is convenient to introduce coordinates $y_1$, $y_2$ and $\rho$ as in Fig.2.
The cross ratio is then written as $\eta=-[(y_1-y_2)^2+\rho^2]/4y_1y_2$.
For the free boundary condition, the correlation must vanish as we go closer
to the boundary:
\beq
\langle\sigma(z_1,\bar z_1)\sigma(z_2,\bar z_2)\rangle_{UHP}
\rightarrow 0,\;\;\mbox{as}\;\;
\rho\rightarrow \infty.
\eeq
Apart from the overall normalisation the coefficients are then determined as
$A=1$ and $B=-1$. 
The scaling law near the boundary is now found to be
\beq
\langle\sigma(z_1,\bar z_1)\sigma(z_2,\bar z_2)\rangle_{UHP}\sim\rho^{-1}.
\eeq
In the case of the fixed boundary condition, the asymptotic behaviour near
the boundary must be
\beq
\langle\sigma(z_1,\bar z_1)\sigma(z_2,\bar z_2)\rangle_{UHP}
\rightarrow 
\langle\sigma(z_1,\bar z_1)\rangle_{UHP}
\langle\sigma(z_2,\bar z_2)\rangle_{UHP},
\eeq
as $\rho\rightarrow \infty$.
The coefficients may be chosen as $A=1$ and $B=1$ to satisfy this condition.
Then, near the boundary we have 
\beq
\langle\sigma(z_1,\bar z_1)\sigma(z_2,\bar z_2)\rangle_{UHP}
\sim (y_1y_2)^{-1/8}.
\eeq
In terms of conformal blocks, the free boundary condition corresponds 
to the process with intermediate $\epsilon$, that is, 
$\langle\sigma\sigma\sigma\sigma\rangle\sim\langle\epsilon\epsilon\rangle$.
The fixed boundary condition corresponds to the identity operator,
$\langle\sigma\sigma\sigma\sigma\rangle\sim\langle II\rangle$.

\begin{figure}
~~~~~~~~~~~~~~~~~~~~~~~~~~~~~~~~~
\epsfxsize=45mm
\epsfysize=70mm
\epsffile{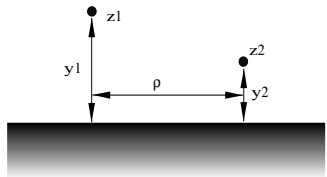}
\vspace{-25mm}
\caption{Parameters $y_1$, $y_2$ and $\rho$.}
\end{figure}


\subsection{Classification of consistent boundary states}

Physical systems described by CFT, such as the Ising model at criticality, 
usually have a finite number of conformally invariant, physically realisable 
boundary states corresponding to various boundary conditions.
For example, in the Ising model there are three physical boundary states 
corresponding to all spins up, down, and free along the boundary.
They are not only conformally invariant but satisfy some extra conditions.
In this section we review Cardy's classification of consistent
boundary states\cite{cardy2}, which uses the modular invariance of partition
functions as the extra information.
In the past several years the study of this method has been expanded 
enormously. 
Generalisations to various rational CFTs, including non-diagonal minimal 
theories\cite{3sp1,3sp2}, superconformal models\cite{nepo},
coset models\cite{coset1,coset2}, have been considered, and algebraic 
understanding of the method\cite{pradisi,runkel,behrend} has also been 
drastically improved. 
We shall not go into these recent developments but describe only the simplest 
diagonal case, following \cite{cardy2,difrancesco}. 

The CFTs we analyse in this subsection are defined on an annulus.
This geometry has a great advantage that the operators on the full plane 
(without boundary) may be employed without modification.
This is due to the fact that in the radial quantisation,
annulus arises as a portion of the full plane bounded by two concentric 
circles.
One may use the conformal transformation 
$w=(T/\pi)\ln z$ and $\zeta=\exp(-2\pi iw/L)$
to map the boundary $z=\bar z$ of the half plane to the two circles
bordering the annulus.
This annulus may also be regarded as a cylinder with length $T$ and circumference
$L$.
On the $\zeta$-plane (annulus), the conformal invariance condition of the 
boundary (\ref{eqn:t=tbar}) becomes
\beq
(L_n-\bar L_{-n})|B\rangle=0.
\label{eqn:conformal}
\eeq
We shall call the boundary states $|B\rangle$ satisfying this condition as
{\em conformally invariant} boundary states.

In ordinary rational conformal theories there is an important set of
conformally invariant boundary states, called {\em Ishibashi} states.
They are defined as
\beq
\vert j\rishi\equiv\sum_M\vert j;M\rangle\otimes U\overline{\vert j;M\rangle},
\eeq
where $j$ is the label for representations, $M$ is the level in the conformal
tower, 
and $U$ is an antiunitary operator which is the product of time reversal and 
complex conjugation.
Ishibashi states are conformally invariant boundary states corresponding 
to conformal towers, and they form a basis spanning the space of 
boundary states. 
An important property of the Ishibashi states is that they diagonalise the
closed string amplitudes and give characters for corresponding representations:
\beq
\lishi i|(q^{1/2})^{L_0+\bar L_0-c/12}|j\rishi
=\delta_{ij}\chi_i(q).
\eeq
These Ishibashi states are not normalisable, as the innerproducts between them
(taking the limit $q\rightarrow 1$ in the expression above) are infinite.

The duality between open and closed string channels imposes a condition on the
boundary states as follows (Fig.3).
Suppose we have boundary conditions $\tilde\alpha$ and $\tilde\beta$ on the
two ends of an open string. 
If these boundary conditions are {\em physical}, chiral representations 
labeled by $i$ appear in the bulk with non-negative integer multiplicities
$n^i_{\tilde\alpha\tilde\beta}$.
The partition function is then the sum of the chiral characters with the 
associated multiplicities, 
\beq 
Z_{\tilde\alpha\tilde\beta}(q)
=\sum_{i}n^i_{\tilde\alpha\tilde\beta}\chi_i(q),
\eeq
where $q=e^{-\pi L/T}$. This is the partition function in the open-string
channel. 
In the closed-string channel, the partition function is nothing but the 
amplitude between two equal-time hypersurfaces,
\beq
Z_{\tilde\alpha\tilde\beta}
=\langle\tilde\alpha\vert
(\tilde q^{1/2})^{L_0+\bar L_0-c/12}
\vert\tilde\beta\rangle,
\eeq
where $\tilde q=e^{-4\pi T/L}$.
Note that the Hamiltonian of our system is 
$2\pi(L_0+\bar L_0-c/12)/L$.
The duality then demands
\beq
\sum_{i}n^i_{\tilde\alpha\tilde\beta}\chi_i(q)
=\langle\tilde\alpha\vert
(\tilde q^{1/2})^{L_0+\bar L_0-c/12}
\vert\tilde\beta\rangle,
\label{eqn:cardy1}
\eeq
which is called Cardy's consistency condition.

By solving this equation, one may find physical boundary conditions and 
express the associated {\em consistent} boundary states as particular linear 
combinations of the Ishibashi states\cite{cardy2}.
Using the modular transformation of the characters
$\chi_i(q)\rightarrow\chi_i(\tilde q)= \sum_jS_{ij}\chi_j(q)$
under $\tau\rightarrow\tilde\tau= -1/\tau$,
the left-hand side of (\ref{eqn:cardy1}) is written as
\beq
\sum_in^i_{\tilde\alpha\tilde\beta}\chi_i(q)
=\sum_{i,j}n^i_{\tilde\alpha\tilde\beta}S_{ij}\chi_j(\tilde q).
\eeq
On the right-hand side, one may expand the boundary states with the Ishibashi
states, as
\beq
\langle\tilde\alpha\vert(\tilde q^{1/2})^{L_0+\bar L_0-c/12}
\vert\tilde\beta\rangle
=\sum_{i,j}\langle\tilde\alpha\vert i\rishi\lishi i\vert
(\tilde q^{1/2})^{L_0+\bar L_0-c/12}\vert j\rishi\lishi j\vert
\tilde\beta\rangle
=\sum_j\langle\tilde\alpha\vert j\rishi\lishi j\vert\tilde\beta\rangle
\chi_j(\tilde q).
\eeq
Equating the coefficients of the character functions on the both sides, we have
\beq
\sum_iS_{ij}n^i_{\tilde\alpha\tilde\beta}=
\langle\tilde\alpha\vert j\rishi\lishi j\vert\tilde\beta\rangle.
\label{eqn:cardy2}
\eeq
Solutions to this equation are found by assuming the existence of a boundary 
state $|\tilde 0\rangle$ satisfying
$n^i_{\tilde 0\tilde\alpha}=n^i_{\tilde\alpha\tilde 0}
=\delta^i_{\tilde\alpha}$ for any boundary condition $\tilde\alpha$.
Letting $\tilde\alpha=\tilde\beta=\tilde 0$ in (\ref{eqn:cardy2}), and using
the positive-definiteness of $S_{0j}$ (which is always the case for unitary
models) we have
\beq
\vert\tilde 0\rangle=\sum_j\sqrt{S_{0j}}\vert j\rishi.
\label{eqn:tilde0}
\eeq
Next, putting $\tilde\alpha=\tilde 0$ and $\tilde\beta\neq\tilde 0$ in
(\ref{eqn:cardy2}) and using the result above, we have
\beq
\vert\tilde\alpha\rangle=\sum_j\frac{S_{\alpha j}}{\sqrt{S_{0j}}}\vert j
\rishi.
\label{eqn:tildealpha}
\eeq
This result (\ref{eqn:tildealpha}) includes the $\tilde\alpha=\tilde 0$ case
(\ref{eqn:tilde0}).

Let us see this result in the case of the critical Ising model.
The operator contents 
$I=\phi_{1,1}=\phi_{2,3}$, $\sigma=\phi_{1,2}=\phi_{2,2}$, 
$\epsilon=\phi_{2,1}=\phi_{1,3}$ of the Kac formula are the identity, 
spin, and energy density operators, respectively.
Substituting the modular S matrix for the Ising model characters (see App.B)
we find consistent boundary states from (\ref{eqn:tildealpha}) as
\bea
|\tilde 0\rangle&=&|\tilde I\rangle
=\frac{1}{\sqrt 2}|I\rangle\!\rangle
+\frac{1}{\sqrt 2}|\epsilon\rangle\!\rangle
+\frac{1}{\sqrt[4]{2}}|\sigma\rangle\!\rangle,\\
&&|\tilde\epsilon\rangle
=\frac{1}{\sqrt 2}|I\rangle\!\rangle
+\frac{1}{\sqrt 2}|\epsilon\rangle\!\rangle
-\frac{1}{\sqrt[4]{2}}|\sigma\rangle\!\rangle,\\
&&|\tilde\sigma\rangle
=|I\rangle\!\rangle-|\epsilon\rangle\!\rangle.
\eea
Since $|\tilde I\rangle$ and $|\tilde \epsilon\rangle$ differ only by the
sign of $|\sigma\rishi$ associated with the spin operator, they are identified
as the fixed boundary conditions ($|\uparrow\rangle$, $|\downarrow\rangle$).
Which is up and which is down is purely a matter of choice. 
The remaining $|\tilde\sigma\rangle$ corresponds to the free boundary 
condition $|F\rangle$.

\begin{figure}
\begin{center}
~
\epsfxsize=45mm
\epsfysize=60mm
\epsffile{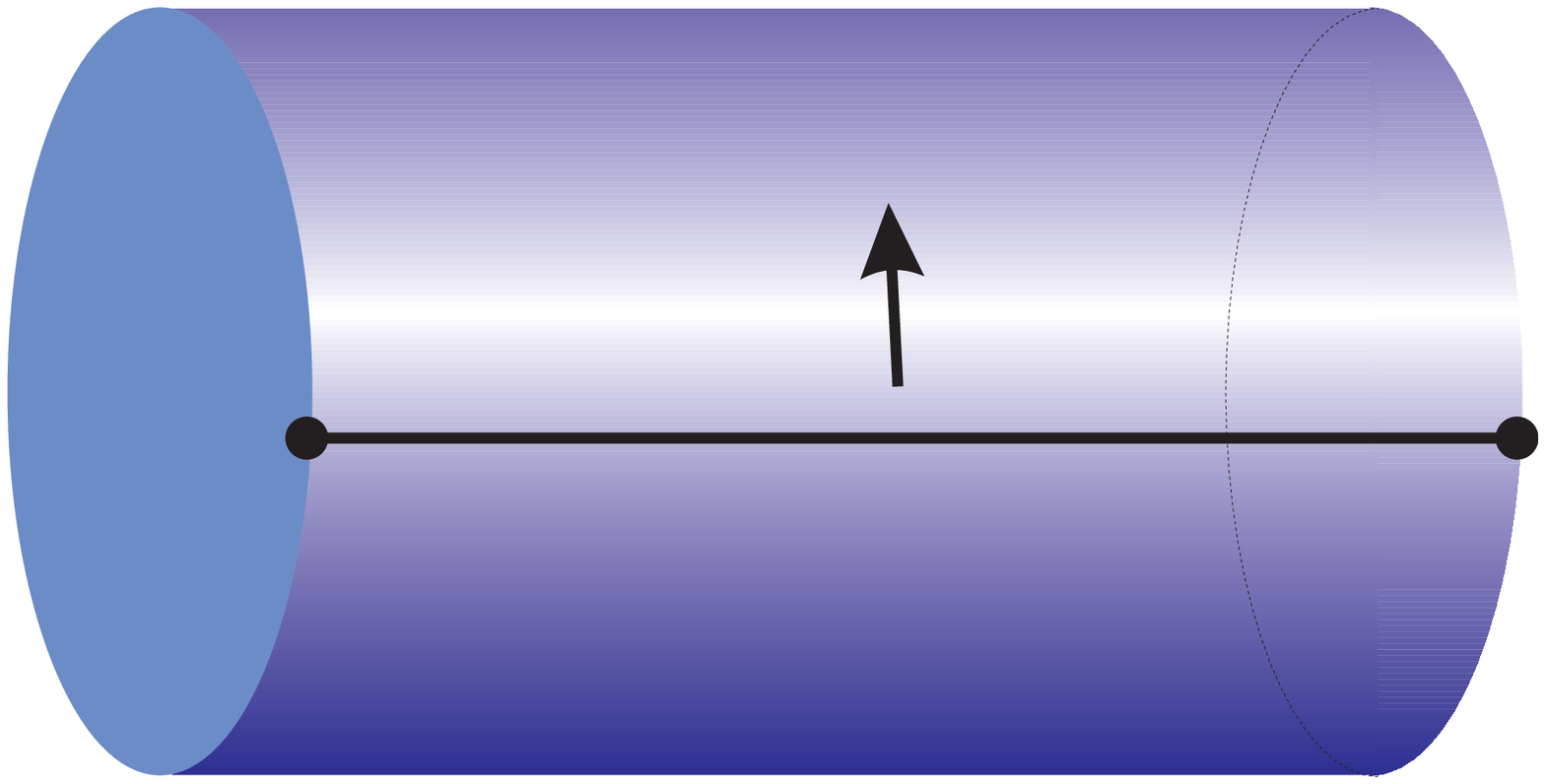}
~~~~~~~~~
\epsfxsize=45mm
\epsfysize=60mm
\epsffile{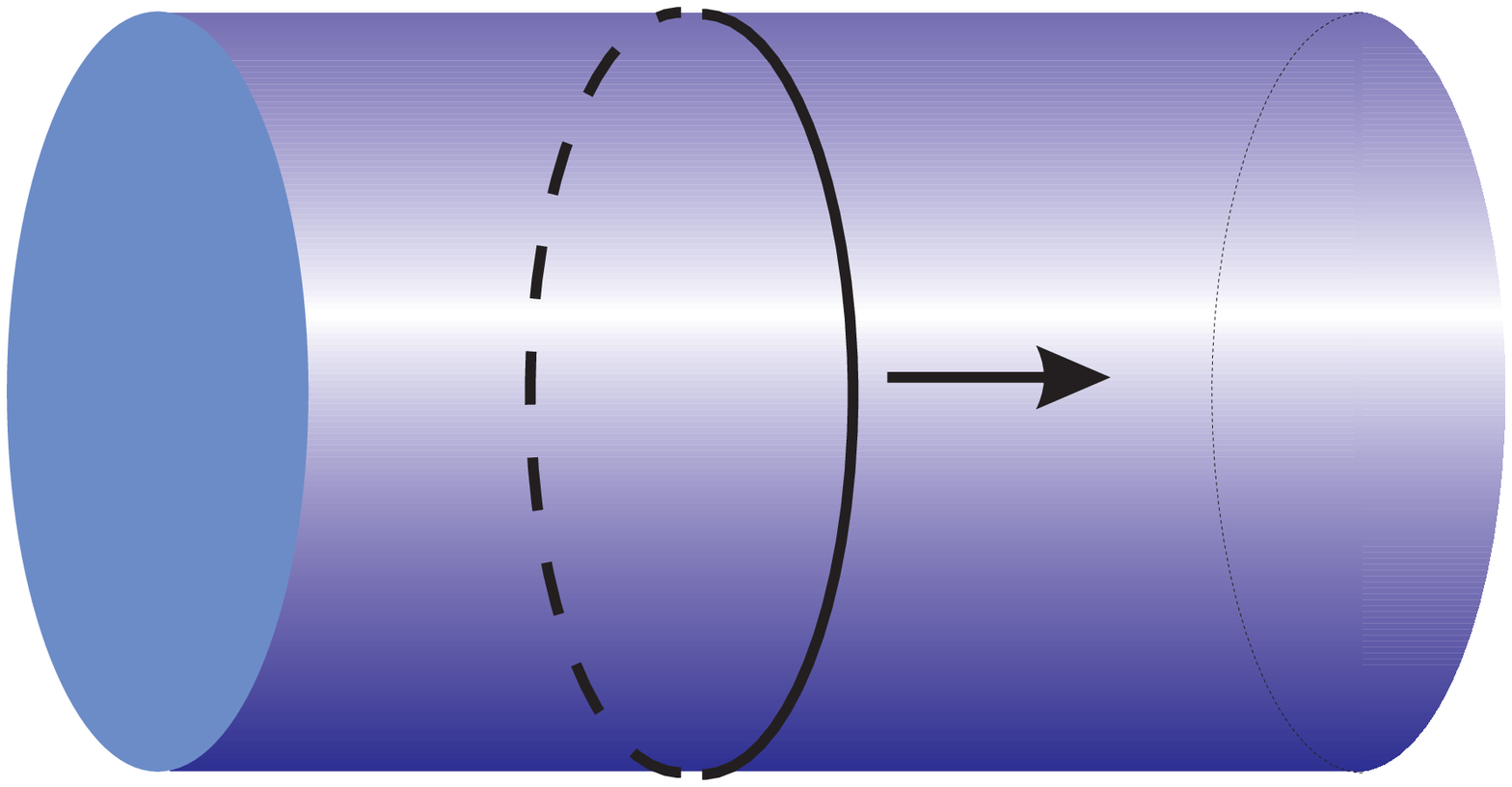}
~\\
(a) Open string channel ~~~~~~~~~~ (b) Closed string channel
\caption{The open-string channel (a) and the closed-string channel (b)
are related by the duality exchanging the directions of time and space. 
The equivalence of the partition functions calculated in each channel 
leads to the constraints (\ref{eqn:cardy1}) on the boundary states.}
\end{center}
\end{figure}


\subsection{Verlinde formula}
One of the most remarkable properties of genus one conformal field theories is
that fusion rules can be determined by the modular transformations of the 
operator contents. 
This is highly non-trivial since fusion is a local property of operators
whereas modular transformations are obviously global.
The relation between fusion and modular transformations is summarised in the 
form of the celebrated Verlinde formula
\cite{verlinde}:
\beq
N_{ij}{}^k=\sum_m\frac{S_{im}S_{jm}\bar S_{mk}}{S_{0m}},
\eeq
where $N_{ij}{}^k$ is the fusion matrix in
$\phi_i\times\phi_j=\sum_k N_{ij}{}^k\phi_k$ and $S_{ij}$ is the modular
S matrix $\chi_i(-1/\tau)=\sum_jS_{ij}\chi_j(\tau)$. 
The index $0$ stands for the vacuum representation.
The proof of this equation is due to the underlying quantum group structure 
applied to genus one  manifold\cite{dij,mooreseiberg}.
See also \cite{gsa} for a more recent review.
Using $SS^\dagger=1$, the above relation may be written in the form
\beq
\sum_k N_{ij}{}^kS_{km}=\frac{S_{im}}{S_{0m}}S_{jm},
\label{eqn:verlinde}
\eeq
meaning that the fusion matrix is diagonalised by the modular S matrix.

In boundary theory, substituting (\ref{eqn:tildealpha}) into the duality 
relation (\ref{eqn:cardy2}) we have
\beq
\sum_iS_{ij}n^i_{\tilde\alpha\tilde\beta}
=S_{\alpha j}S_{\beta j}/S_{0j}.
\eeq
Comparing this with (\ref{eqn:verlinde}), it is concluded that\cite{cardy2}
\beq
N_{ij}{}^k=n^k_{ij},
\eeq
that is, the multiplicity of the representations appearing in the bulk is
identical to the fusion coefficient for the operators associated to the 
boundary states.


\subsection{Boundary operators and critical percolation}

The mirroring method reviewed in Sec.2.2 naturally introduces another 
important object in boundary CFT, called a boundary 
operator\cite{cardylewellen},
which is defined through the OPE of two bulk operators, 
one on the upper and the other on the lower half plane, 
\beq
\phi_i(z)\phi_j(z^*)\sim\sum_k(z-z^*)^{(h_k-h_i-h_j)}N_{ij}{}^k\phi_{B,k}(x),
\eeq
where $x=(z+z^*)/2$.
The $\phi_{B,k}(x)$ are the boundary operators which reside {\em on}
the boundary.
Such a boundary operator may change the boundary condition when inserted. 
The operator changing the boundary state from $\tilde\alpha$
to $\tilde\beta$ is denoted as $\phi_{(\tilde\alpha\tilde\beta)}$. 

The critical bond percolation problem in statistical physics is an example 
which is solved using boundary operators.
Since this is often discussed in the context of logarithmic CFTs, we shall 
review it here briefly\cite{carperc,difrancesco,henkel}. 

The problem we want to solve is defined as follows.
We consider a lattice of horizontal length $a\ell$, vertical length $b\ell$ 
and spacing $\ell$, and set electrodes on the left and right sides of the 
lattice.
We start placing conducting needles randomly on the grid, and observe if 
electric current can run between the two electrodes (horizontal percolation).
Obviously, when we put no needle on the lattice there is no way the current
can run through, and when all the grids are filled with conducting needles the
percolation has readily been achieved. 
Thus there must be some occupation probability $p$ between $0$ and $1$ which
is barely sufficient to achieve the percolation. 
We may take the thermodynamic limit of this system (letting 
$\ell\rightarrow 0$ and $a\ell$, $b\ell$ fixed). 
Then there is a critical occupation probability $p_c$ such
that the horizontal percolation probability $\pi_h$ is one for $p>p_c$ and
$\pi_h=0$ for $p<p_c$. 
The system at $p=p_c$ is called the critical bond percolation.
At $p=p_c$, $\pi_h$ still depends on the aspect ratio $r=a/b$.
Our problem is to find $\pi_h$ as a function of $r$.

This percolation problem is translated into $Q\rightarrow 1$
limit of the Q-state Potts model in two dimension\cite{kasteleyn,wu}.
The interaction energy of the Q-state Potts model is 
$J\sum_{\langle ij\rangle}\delta_{\sigma_i\sigma_j}$, where the sum is over 
nearest-neighbours and the indices $i$ and $j$ take one of the Q states.
The partition function of the Q-state Potts model is then
\beq
Z=\sum_{\mbox{\scriptsize config}}\prod_{\langle ij\rangle}
\exp\left\{-\beta J\delta_{\sigma_i\sigma_j}\right\}.
\eeq
Defining $1/(1-p)=e^{-\beta J}$, one may rewrite this partition function as
\beq
Z=\sum_{\mbox{\scriptsize activation}}p^B(1-p)^{b-B}Q^{N_c},
\eeq
where $b$ is the number of bonds, $B$ is the number of activated bonds 
(needles), and $N_c$ is the number of disjoint clusters. 
The Q-state Potts model has been mapped into a system with
active bonds of Q possible colours (probability $=p$) and inert bonds
(probability$=1-p$).
It is now obvious that the critical percolation is realised by taking
$Q\rightarrow 1$ limit of this system. 

The relation between the Q-state Potts models and the ${\cal M}(m+1,m)$ 
unitary 
minimal series ($m=3$ is the Ising model) is well established for $Q=2,3,4$.
The correspondence is given by $Q=4\cos^2(\pi/(m+1))$ for these models.
Extrapolating this formula for arbitrary value of $m$, the bond percolation 
problem then correspond to the minimal model of ${\cal M}(3,2)$, 
whose central charge is $c=0$.
Using this correspondence, the percolation problem is described in a boundary 
CFT language as follows.
Let us consider a rectangle with a pair of opposing sides having free 
boundary condition $\tilde f$ (Fig.4). 
The remaining two sides have fixed boundary conditions $\tilde\alpha$ and
$\tilde\beta$, with fixed `colours' out of the Q possible colours of the 
activated bonds. 
Such a configuration is realised by inserting four boundary (changing)
operators at the four corners of the rectangle.
The boundary operator $\phi_{(\tilde\alpha\tilde f)}$, which changes a fixed 
boundary condition to the free boundary condition, is identified as
$\phi_{2,1}$ by analogy with the Q-state Potts models of $Q=2,3,4$.

\begin{figure}
\unitlength 1mm
\begin{center}
\begin{picture}(80,50)
\multiput(0,0)(2,0){30}{\line(1,0){0.5}}
\put(0,0){\line(0,1){40}}
\put(60,0){\line(0,1){40}}
\multiput(0,40)(2,0){30}{\line(1,0){0.5}}
\put(-5,20){$\tilde\alpha$}
\put(62,20){$\tilde\beta$}
\put(20,-5){Free b.c.}
\put(20,45){Free b.c.}
\put(-5,-5){$\phi_{(\tilde f\tilde\alpha)}(x_1)$}
\put(-5,45){$\phi_{(\tilde\alpha\tilde f)}(x_4)$}
\put(62,-5){$\phi_{(\tilde\beta\tilde f)}(x_2)$}
\put(62,45){$\phi_{(\tilde f\tilde\beta)}(x_3)$}
\end{picture}
\end{center}
\vspace{10mm}
\caption{Horizontal bond percolation is modelled by a rectangle with
the free boundary condition on the top and bottom sides, and fixed boundary 
conditions on the left and right sides.} 
\end{figure}
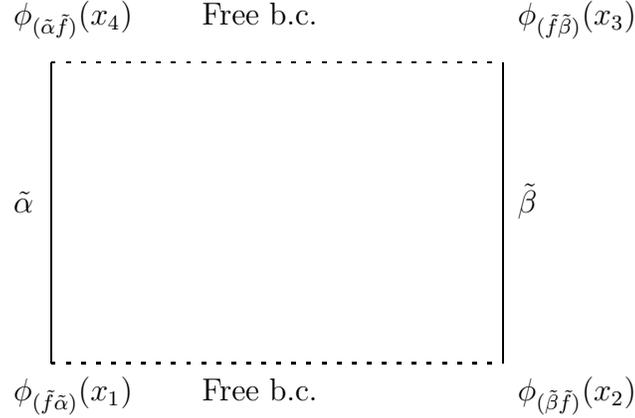

The crossing probability $\pi_h$ is obtained by calculating the partition
functions for the configurations
\unitlength 1mm
\beq
\left(
\begin{picture}(30,15)
\put(5,-7){\line(0,1){15}}
\put(25,-7){\line(0,1){15}}
\put(5,3){\line(1,0){5}}
\put(10,3){\line(0,-1){5}}
\put(10,-2){\line(1,0){5}}
\put(15,-2){\line(0,-1){5}}
\put(15,-7){\line(1,0){5}}
\put(20,-7){\line(0,1){5}}
\put(20,-2){\line(1,0){5}}
\put(2,0.5){$\tilde\alpha$}
\put(26,0.5){$\tilde\alpha$}
\put(15,-1){$\tilde\alpha$}
\end{picture}
\right)+\left(
\begin{picture}(30,15)
\put(5,-7){\line(0,1){15}}
\put(25,-7){\line(0,1){15}}
\put(5,3){\line(1,0){5}}
\put(10,3){\line(0,-1){5}}
\put(10,-2){\line(1,0){5}}
\put(15,-2){\line(0,-1){5}}
\put(15,-7){\line(1,0){5}}
\put(15,3){\line(1,0){5}}
\put(20,-2){\line(0,1){5}}
\put(20,-2){\line(1,0){5}}
\put(2,0.5){$\tilde\alpha$}
\put(26,0.5){$\tilde\alpha$}
\put(11,-6){$\tilde\alpha$}
\put(16,4){$\tilde\alpha$}
\end{picture}
\right)-\left(
\begin{picture}(30,15)
\put(5,-7){\line(0,1){15}}
\put(25,-7){\line(0,1){15}}
\put(5,3){\line(1,0){5}}
\put(10,3){\line(0,-1){5}}
\put(10,-2){\line(1,0){5}}
\put(15,-2){\line(0,-1){5}}
\put(15,-7){\line(1,0){5}}
\put(15,3){\line(1,0){5}}
\put(20,-2){\line(0,1){5}}
\put(20,-2){\line(1,0){5}}
\put(2,0.5){$\tilde\alpha$}
\put(26,0.5){$\tilde\beta$}
\put(11,-6){$\tilde\alpha$}
\put(16,4){$\tilde\beta$}
\end{picture}
\right)
\label{eqn:percgraph}
\eeq
and taking the limit $Q\rightarrow 1$ (that is, 
$\tilde\alpha=\tilde\beta$) afterwards.
Note that in this limit the last two terms cancel and only the first one
(realising the percolation) remains.
If we write the partition function for the configurations with the boundary 
condition $\tilde\alpha$ on the left and $\tilde\beta$ on the right as 
$Z_{\tilde\alpha\tilde\beta}$, the crossing probability is given by
\beq
\pi_h=\lim_{Q\rightarrow 1}(Z_{\tilde\alpha\tilde\alpha}
-Z_{\tilde\alpha\tilde\beta}),
\eeq
since the first two terms of the graphical representation (\ref{eqn:percgraph})
are $Z_{\tilde\alpha\tilde\alpha}$ and the last one is
$Z_{\tilde\alpha\tilde\beta}$.
These partition functions for particular boundary conditions are given
by the four-point functions of the boundary operators.
Then up to a multiplicative constant we have 
\bea
\pi_h
&\sim&\langle\phi_{(\tilde f\tilde\alpha)}(x_1)
\phi_{(\tilde\alpha\tilde f)}(x_2)
\phi_{(\tilde f\tilde \beta)}(x_3)
\phi_{(\tilde\beta\tilde f)}(x_4)\rangle_{Q=1}\nonumber\\
&=&\langle\phi_{12}(x_1)\phi_{12}(x_2)
\phi_{12}(x_3)\phi_{12}(x_4)\rangle_{Q=1}.
\eea
The four-point function 
$\langle\phi_{12}(x_1)\phi_{12}(x_2)\phi_{12}(x_3)\phi_{12}(x_4)\rangle$
is found by solving a second order ordinary differential equation.
Introducing the cross ratio $\eta=[(z_4-z_3)(z_2-z_1)]/[(z_3-z_1)(z_4-z_2)]$,
where $z_i$ are the coordinates after the Schwartz-Christoffel transformation
$x_i\rightarrow z_i$ mapping the interior of the rectangle to the upper
half plane, the differential equation for $\pi_h(\eta)=g(\eta)$ is 
\beq
\eta(1-\eta)\frac{d^2g(\eta)}{d\eta^2}
+\frac{2(1-2\eta)}{3}\frac{dg(\eta)}{d\eta}=0.
\label{eqn:percde}
\eeq
This differential equation has two independent solutions. 
One of them is $g(\eta)=const$. The other is
\beq
g(\eta)=\eta^{1/3}F(\frac 13,\frac 23,\frac 43;\eta).
\label{eqn:geta}
\eeq
The crossing probability $\pi_h$ is a linear combination of the two solutions.
The coefficients are determined by demanding $\pi_h\rightarrow 1$ when the 
rectangle is infinitely narrow and $\pi_h\rightarrow 0$ when it is infinitely 
wide.
The solution is then
\beq
\pi_h(\eta)
=\frac{3\Gamma(2/3)}{\Gamma(1/3)^2}\eta^{1/3}
F(\frac 13,\frac 23,\frac 43;\eta).
\eeq
This analytic result is compared with numerical calculations and exhibits 
excellent agreement\cite{carperc,lang,aizenman}. 
Although the extrapolation of the $Q\neq 1$ results to $Q\rightarrow 1$ may 
seem somewhat speculative, this agreement justifies the method of the
analysis as well as the underlying concepts such as conformal invariance and 
boundary operators.

Finally we emphasise that this CFT at $c=0$ is {\em not} the minimal model of
${\cal M}(3,2)$, which consists only of the identity operator.
Recall that the differential equation (\ref{eqn:percde}) has two independent 
solutions, one corresponds to the conformal block $\phi_{1,1}$ and the 
other to $\phi_{1,3}$. 
Obviously, the former solution is the constant and the latter is 
(\ref{eqn:geta}).
If we were dealing with ${\cal M}(3,2)$ minimal model, the solution 
(\ref{eqn:geta}) 
should have been discarded since it is associated to the operator outside the
Kac table.
Hence the percolation problem must be considered in the framework of 
a CFT with extended conformal grid, possibly to 
${\cal M}(9,6)$\cite{mflohr,henkel}.
From this example we may expect the existence of bona-fide CFTs which are
not minimal models but something that should be called
`next-to-minimal' models, which may well include logarithmic 
operators\cite{communalex}. 
The existence of such statistical models is one of the biggest reasons for
studying logarithmic CFTs with boundary.
\newpage

\section{Logarithmic CFT with boundary}


In this section we discuss logarithmic conformal field theories with boundary.
Probably the biggest motivation for the study of such theories is the 
existence of a number of statistical systems which are conjectured to be 
modelled by logarithmic CFTs, where the finite
size effect may not be negligible if the system has a boundary. 
There are also examples in which boundary operators play an essential role, 
as is mentioned at the end of the last section.  
As logarithmic CFTs are in general non unitary, being far from
boundary does not guarantee the irrelevance of the boundary effect.
Once we have a boundary, its effect may change the system globally.
Therefore the study of boundary effect seems to be extremely important for
a proper CFT description of non-unitary systems. 
Also, in string theory context, there are several examples such as 
D-brane recoil\cite{dbrecoil1,dbrecoil2} and dimensional 
reduction\cite{dimred}, where 
logarithmic operators are claimed to play important roles. 
These are less established theories than the statistical model examples,
and certainly more investigation is required.
If they turn out to be correct descriptions, the boundary theory of such 
systems will be important since open strings and branes necessarily 
involve world sheet boundaries.

When we try to apply standard techniques of boundary CFT to logarithmic cases,
several unusual features arise.
In the following we shall see such results for the $c=-2$ logarithmic CFT.
In the next subsection we calculate correlation functions in the presence of
boundary and argue that logarithmic behaviour is unavoidable either near 
or away from the boundary\cite{koganwheater}.
In Subsec.3.2 we review free-field representation of the $c=-2$ model,
following \cite{local,curious,sf}.
Even on genus one manifold without boundary, the logarithmic CFT at $c=-2$ is 
somewhat pathological, in that the fusion matrix is not diagonalisable and 
thus the Verlinde formula fails.
We shall review this in Sec. 3.3. 
As Cardy's method depends largely on the modular properties of CFT, 
the classification of boundary states in logarithmic CFT does not go
straightforwardly. 
However, we can still find `good' boundary states with consistent modular
properties, as is discussed in Sec. 3.4.  


\subsection{Boundary correlation functions in LCFT}
The CFT at $c=c_{2,1}=-2$ is one of the logarithmic CFTs that have been
studied most intensively and are so far best understood. 
This was used by Gurarie\cite{gurarie} to discuss the importance of 
logarithmic operators, and it is also claimed that certain universality 
classes of two-dimensional statistical models (such as critical polymers in 
the dense phase\cite{saleur}) are described by this theory.
Let us start by discussing operator contents of this theory, assuming that
they are given by extending the Virasoro minimal models. 
Of course the Kac determinant must be reconsidered to get a proper 
formula consistent with the non-trivial Jordan cell\cite{mflohr,flohrsv}. 
However we shall adopt the original formula here, as it still contains
some truth at least for the pre-logarithmic part. 

Usually the operator content of the ${\cal M}(p,p')$ minimal model 
(we assume $p>p'$) 
is restricted to $\phi_{r,s}$ such that $0<r<p'$, $0<s<p$ (and also $pr>p's$
to avoid the double counting of identical operators).
Since the Kac table for $(p,p')=(2,1)$ is empty, the border of the grid must 
be extended in order to have a non-trivial theory.
Note that the Kac formula of conformal dimension $h_{r,s}$ for 
${\cal M}(p,p')$ minimal model,
\beq
h_{r,s}=\frac{(pr-p's)^2-(p-p')^2}{4pp'},
\eeq
is invariant for the ``rescaling'' $p\rightarrow lp$ and $p'\rightarrow lp'$
for some natural number $l$.
The table of conformal dimensions for the extended ${\cal M}(2,1)$ minimal 
model is

\begin{center}
\begin{tabular}{c|cccccccccccc} 
\hline
&\multicolumn{12}{c}{$s$}\\
\cline{2-13}
$r$&$1$&$2$&$3$&$4$&$5$&$6$&$7$&$8$&$9$&$10$&$11$&$12$\\
\hline\hline
$1$&$0$&$-1/8$&$0$&$3/8$&$1$&$15/8$&$3$&$35/8$&$6$&$63/8$&$10$&$\cdots$\\
$2$&$1$&$3/8$&$0$&$-1/8$&$0$&$3/8$&$1$&$15/8$&$3$&$35/8$&$6$&${}$\\
$3$&$3$&$15/8$&$1$&$3/8$&$0$&$-1/8$&$0$&$3/8$&$1$&$15/8$&$3$&${}$\\
$4$&$6$&$35/8$&$3$&$15/8$&$1$&$3/8$&$0$&$-1/8$&$0$&$3/8$&$1$&${}$\\
$5$&$10$&$63/8$&$6$&$35/8$&$3$&$15/8$&$1$&$3/8$&$0$&$-1/8$&$0$&${}$\\
$6$&$\cdots$&${}$&${}$&${}$&${}$&${}$&${}$&${}$&${}$&${}$&${}$&$\cdots$\\
\hline
\end{tabular}
\end{center}
We shall restrict the contents to $0<r<3$, $0<s<6$, that is, we consider the
`next-to-minimal' model ${\cal M}(6,3)$. 
It has been shown by a free-field representation that these operators indeed
close under the fusion rule. 

The operators $\phi_{1,1}$, $\phi_{1,3}$, and $\phi_{1,5}$ belong to a
degenerate `logarithmic' block and are expected to be governed by unusual 
Ward identities\cite{flohrsv}.
However, $\mu=\phi_{1,2}$ and $\nu=\phi_{1,4}$ are normal (pre-logarithmic) 
operators and therefore satisfy ordinary conformal Ward identities.
This enables us to calculate boundary $n$-point functions involving $\mu$ and 
$\nu$\cite{koganwheater,ma-rouhani}, using the mirroring method described in 
Subsec.2.2.
Let us see this in a simple example, boundary $2$-point function of the spin
$-1/8$ operators $\langle\mu\mu\rangle_B$\cite{koganwheater}.
As the boundary $n$-point functions are equivalent to {\em chiral} $2n$-point 
functions, in our case we have
\beq
\langle\mu(z_1,\bar z_1)\mu(z_2,\bar z_2)\rangle_B
=\langle\mu(z_1)\mu(z_2)\mu(z_1^*)\mu(z_2^*)\rangle_{chiral}.
\eeq
Since $\mu=\phi_{1,2}$, this chiral $4$-point function satisfies a second order
differential equation, which reduces to a hypergeometric differential equation
after M\"obius transformations.
The solution is easily found as
\bea
&&\langle\mu(z_1,\bar z_1)\mu(z_2,\bar z_2)\rangle_B\nonumber\\
&&=|z_1-z_2^*|^{1/2}z^{1/4}(1-z)^{1/4}
\left\{AF(\frac 12, \frac 12, 1;z)
+BF(\frac 12, \frac 12, 1;1-z)\right\},
\eea
where $z$ is the cross ratio,
\beq
z=\frac{(z_1-z_2)(z_1^*-z_2^*)}{(z_1-z_2^*)(z_1^*-z_2)}
=\frac{|z_1-z_2|^2}{|z_1-z_2^*|^2},
\eeq
and $A$ and $B$ are constants to be determined by boundary conditions.
Note that this is a single-valued function since $z$ is always real and
$0<z<1$.
The function $F(1/2,1/2,1;z)$ is a hypergeometric function of Gaussian type 
and reduces to the complete elliptic integral
\beq
\frac{\pi}{2}F(\frac 12, \frac 12, 1;k^2)=K(k)
=\int_{0}^{\pi/2}\frac{d\theta}{\sqrt{1-k^2\sin^2\theta}}.
\eeq

If the points $z_1$ and $z_2$ are away from the boundary but the separation
is kept fixed, we have $z\rightarrow 0$ and then
\beq
\langle\mu(z_1,\bar z_1)\mu(z_2,\bar z_2)\rangle_B\rightarrow
A|z_1-z_2|^{1/2}+2B|z_1-z_2|^{1/2}\ln |z_1-z_2|.
\eeq
The first term is the same as the bulk $2$-point function. 
Hence, we may let $B=0$ if we want to recover the bulk result by letting $z_1$ 
and $z_2$ away from the boundary. 
However, there is no physical motivation to do so because our theory is not
unitary and the $2$-point function grows with the separation. 
That is, being away from the boundary does not guarantee the negligibility
of the boundary effect.

If the two operators are close to the boundary, we have $z\rightarrow 1$
and then $F(\frac 12, \frac 12, 1;1-z)\rightarrow 1$.
The correlation function is dominated by the first term,
\beq
\langle\mu(z_1,\bar z_1)\mu(z_2,\bar z_2)\rangle_B\sim
2A|z_1-z_1^*|^{1/2}\ln\frac{|z_1-z_1^*|}{|z_1-z_2|},
\eeq
displaying a logarithmic behaviour.
We conclude that the $2$-point function of pre-logarithmic operators exhibit
logarithmic divergence either away from or close to the boundary.


\subsection{Triplet $c=-2$ model in free-field representation}

Extension of the Kac table discussed in the previous subsection allows
us to use differential equations for correlation functions and in some 
situations (when the correlators involve non-logarithmic operators) we can 
find explicit form of correlation functions with or without boundary. 
Such a method is quite powerful and is apparently a correct approach at least 
for some cases.
For example, in the percolation problem reviewed in Subsec.1.5, we have used a
solution of a differential equation associated to an operator outside the
Kac table, and the result is supported by a remarkable agreement with numerical
calculations. 

However, it is desirable to start a theory from a more solid ground,
as we do not know to what extent the results of the unitary minimal models
may be generalised to logarithmic cases without modification.
In $c=-2$ logarithmic CFT, a construction based on a free-field
representation does exist, and the so-called triplet model has been shown to
be realised by symplectic fermions. 
In this subsection we review the free-field representation of this model
following \cite{curious,sf} and collect basic results needed for following
discussions.
For more complete descriptions readers are referred to the original 
papers\cite{curious,sf}. 
See also \cite{rational,local,eholzer} and the lecture note by 
Gaberdiel\cite{mgaberdiel}.  

Symplectic fermions have the same central charge $c=-2$ as the simple ghost 
system whose action is
\beq
S=\frac{1}{\pi}\int d^2z (\eta\bar\partial\xi+\bar\eta\partial\bar\xi),
\label{eqn:action}
\eeq
where $\eta$ and $\xi$ are fermionic ghosts with conformal
dimensions $h_\eta=1$ and $h_\xi=0$.
The operator products are
\beq
\eta(z)\xi(w)\sim\xi(z)\eta(w)\sim1/(z-w),
\eeq
reflecting the Grassmannian nature of the operators.
Symplectic fermions are introduced as operators of conformal dimensions
$h_{\chi^{\pm}}=1$, defined as
\beq
\chi^+\equiv\eta,\;\; \chi^-\equiv\partial\xi.
\eeq
The mode expansions,
\beq
\chi^{\pm}(z)=\sum_{k\in Z}\chi^{\pm}_kz^{-k-1},
\eeq
define the mode operators $\chi^{\pm}_k$ with anti-commutation relations,
\beq
\{\chi^\alpha_m,\chi^\beta_n\}=md^{\alpha\beta}\delta_{m+n},
\eeq
where $d^{\alpha\beta}$ is antisymmetric and $d^{\pm\mp}=\pm 1$.
These symplectic fermions differ from the $\eta$-$\xi$ simple ghost
system by the zero-mode of $\xi$.
The absence of $\xi_0$ enhances the symmetry and realises the triplet model 
at $c=-2$.

Orbifold structure is endowed by considering twisted sectors as well as the
untwisted sector\cite{saleur,curious,sf}. 
The twisted sectors are built on the vacuum by operating with a twisting field
$\sigma_{k/N}$, and the resulting theory becomes $Z_N$ invariant.
It is argued that the $Z_N$ orbifold model constructed like this has a 
W-algebra
of type ${\cal W}(2,3,N(N+1)/2, N(N+1)/2)$\cite{eholzer}.
In the case of $N=2$ the model possesses ${\cal W}(2,3,3,3)$ symmetry which is
generated by the stress tensor
\beq
T(z)=:\chi^-(z)\chi^+(z):+\lambda(\lambda-1)/2z^2,
\eeq
where $\lambda=0$ for the untwisted and $\lambda=1/2$ for the twisted sector,
and a triplet of W-fields with conformal dimension 3,
\bea
&&W^0=-\frac 12 (:\del\chi^+(z)\chi^-(z):+:\del\chi^-(z)\chi^+(z):),\nonumber\\
&&W^\pm=:\del\chi^\pm(z)\chi^\pm(z):.
\eea
Virasoro operators and $W$-mode operators are found from these as
\beq
L_n=\frac 12 d_{\alpha\beta}\sum_{m\in Z+\lambda}
:\chi^\alpha_m\chi^\beta_{n-m}:
+\frac{\lambda(\lambda-1)}{2}\delta_{n0},
\label{eqn:virop}
\eeq
and
\bea
W^0_n&=&-\frac 12\sum_{j\in Z+\lambda}j
\left\{:\chi^+_{n-j}\chi^-_j:+:\chi^-_{n-j}\chi^+_j:\right\},\nonumber\\
W^\pm_n&=&\sum_{j\in Z+\lambda}j\chi^\pm_{n-j}\chi^\pm_j.
\label{eqn:wmode}
\eea
In the following we only consider the $N=2$ case which is called the triplet
model for an obvious reason.

The representations with conformal weights $-1/8$, $0$, $3/8$ and $1$ 
expected from the Kac table at $c_{2,1}=-2$ are associated to the Fock space 
representations of the symplectic fermions as 
follows\cite{local,mgaberdiel,sf}.
The ground state of the twisted sector which is obtained by operating
with $\sigma_{1/2}$ on the vacuum is denoted by $\mu$. 
This $\mu$ has conformal weight $-1/8$, and then the singlet representation 
${\cal V}_{-1/8}$ is defined as states built on $\mu$.
The doublet of the states $\nu^\pm\equiv\chi^\pm_{-1/2}\mu$
has conformal dimension $3/8$ and this is the highest-weight states
of the doublet representation ${\cal V}_{3/8}$. 
Since ${\cal V}_{3/8}$ corresponds to $\phi_{1,4}$ or $\phi_{2,2}$ in the Kac 
table,
the correlation functions involving ${\cal V}_{3/8}$ satisfy fourth order
differential equations, whose solutions represent two conformal blocks for 
$\nu^+$ and $\nu^-$ each.
The representations belonging to the untwisted sector are more complicated.
Let $\omega$ be a state annihilated by operations with $\chi^\pm_{n>0}$.
Then there are four ground states, $\omega$, $\theta^\pm=-\chi^\pm_0\omega$,
and $\Omega=\chi^-_0\chi^+_0\omega=L_0\omega$. As $\Omega$ is annihilated by
further operations with zero modes, it is identified as the M\"obius invariant
vacuum.
The irreducible vacuum representation ${\cal V}_0$ is built on the ground
state $\Omega$. Similarly, the irreducible doublet representation ${\cal V}_1$ 
at $h=1$ is built on the doublet $\psi^\pm=\chi^\pm_{-1}\Omega$.

These four representations ${\cal V}_{-1/8}$, ${\cal V}_{3/8}$,
${\cal V}_0$ and ${\cal V}_1$ do {\em not} close under the fusion and thus
extra representations are needed. 
The `reducible but indecomposable' representation ${\cal R}_0$ is obtained by
extending the vacuum ${\cal V}_0$ to include $\omega$, $L_{-1}\omega$ and 
$W^a_{-1}\omega$ as well as $\Omega$.
The two bosonic ground states $\Omega$ and $\omega$  span a two-dimensional
Jordan cell on the action of $L_0$, forming a `logarithmic pair.'
The representation ${\cal R}_1$ is obtained likewise, by extending ${\cal V}_1$
to include $\phi^\pm=\chi^\pm_{-1}\omega$ and $\xi^\pm=-L_1\phi^\pm$ in 
addition to $\psi^\pm=\chi^\pm_{-1}\Omega$.
The doublet states $\psi^\pm$ and $\phi^\pm$ form a logarithmic pair at $h=1$.
The four representations ${\cal V}_{-1/8}$, ${\cal V}_{3/8}$, ${\cal R}_0$ and 
${\cal R}_1$ close under the fusion.

The representations in the triplet model are summarised as follows:\\\\
\underbar{Twisted sector}
\begin{displaymath}
\begin{array}{ccc}
\mu&\rightarrow&{\cal V}_{-1/8}\\
\nu^\pm\equiv\chi^\pm_{-1/2}\mu&\rightarrow&{\cal V}_{3/8}
\end{array}
\end{displaymath}
\underbar{Untwisted sector}
\begin{displaymath}
\begin{array}{ccc}
\Omega&\rightarrow&{\cal V}_{0}\\
\psi^\pm\equiv\chi^\pm_{-1}\Omega&\rightarrow&{\cal V}_{1}
\end{array}
\end{displaymath}
\begin{displaymath}
\begin{array}{ccc}
\omega,\;\; \Omega,\;\; L_{-1}\omega,\;\; 
W^a_{-1}\omega&\rightarrow&{\cal R}_0\\
\phi^\pm\equiv\chi^\pm_{-1}\omega,\;\; \psi^\pm,\;\; 
\xi^\pm\equiv-L_1\phi^\pm&\rightarrow
&{\cal R}_1
\end{array}
\end{displaymath}


\subsection{Fusion rules and modular invariants}
The fusion rules of the $c=-2$ triplet model are calculated both in the 
algebraic method\cite{rational,mgaberdiel} and the free-field method\cite{sf} 
to be
\beq
\begin{array}{lll}
{\cal R}_i\times{\cal R}_j & =2{\cal R}_0+2{\cal R}_1 & i,j=0,1,\\
{\cal R}_i\times{\cal V}_j & ={\cal R}_0 & (i,j)=(0,0),(1,1),\\
& ={\cal R}_1 & (i,j)=(0,1),(1,0),\\ 
& =2{\cal V}_{-1/8}+2{\cal V}_{3/8} 
& i=0,1; j=-\frac 18,\frac 38,\\
{\cal V}_i\times{\cal V}_j & ={\cal V}_0 & (i,j)=(0,0),(1,1),\\
& ={\cal V}_1 & (i,j)=(0,1),(1,0),\\ 
& ={\cal V}_{-1/8} & (i,j)=(0,-\frac 18), (1,\frac 38),\\
& ={\cal V}_{3/8} & (i,j)=(1,-\frac 18), (0,\frac 38),\\
& ={\cal R}_0 & (i,j)=(-\frac 18,-\frac 18), (\frac 38,\frac 38),\\
& ={\cal R}_1 & (i,j)=(-\frac 18,\frac 38),(\frac 38,-\frac 18).
\end{array}
\label{eqn:fusion}
\eeq
Since the four representations ${\cal R}_0$, ${\cal R}_1$, ${\cal V}_{-1/8}$
and ${\cal V}_{3/8}$ close under this fusion rule, the triplet model can be 
regarded
as a rational conformal field theory, with a weakened definition of 
rationality\cite{rational}.

What is unusual about this fusion rule is that it is not diagonalisable. 
This is in a sharp contrast with ordinary rational theories, where fusion
matrices are always diagonalised with modular S matrices through the Verlinde 
formula.
As is expected from the failure of the Verlinde formula, the characters of
the triplet model are quite unusual as well.
They are calculated in \cite{rational,sf,flohrmod1,flohrmod2} as
\begin{eqnarray}
&&\chi_{{\cal V}_0}(\tau)
=\frac{\Theta_{1,2}(\tau)}{2\eta(\tau)}+\frac 12 \eta(\tau)^2,\nonumber\\
&&\chi_{{\cal V}_1}(\tau)
=\frac{\Theta_{1,2}(\tau)}{2\eta(\tau)}-\frac 12 \eta(\tau)^2,\nonumber\\
&&\chi_{{\cal V}_{-1/8}}(\tau)
=\frac{\Theta_{0,2}(\tau)}{\eta(\tau)},\nonumber\\
&&\chi_{{\cal V}_{3/8}}(\tau)=\frac{\Theta_{2,2}(\tau)}{\eta(\tau)},\nonumber\\
&&\chi_{{\cal R}_0}(\tau)
=\chi_{{\cal R}_1}(\tau)=\frac{2\Theta_{1,2}(\tau)}{\eta(\tau)},
\end{eqnarray}
where $\Theta_{k,l}(\tau)$ and $\eta(\tau)$ are Jacobi theta functions and
Dedekind eta function, respectively (see App.B for definitions).
Note that these character functions are not independent,
$\chi_{{\cal R}_0}=\chi_{{\cal R}_1}=2\chi_{{\cal V}_0}+2\chi_{{\cal V}_1}$.
The mutual dependence of the character functions is reminiscent of the minimal
Virasoro theories with extra symmetry (such as the three-state Potts model).
What is more pathological is the modular transformation property of these
characters.
Since $\eta(\tau)^2\rightarrow\eta(\tilde\tau)^2=-i\tau\eta(\tau)^2$ as 
$\tau\rightarrow\tilde\tau=-1/\tau$,
the character functions do not transform each other linearly under the modular 
transformation.

Nevertheless, we may construct a modular-invariant partition function from 
these character functions\cite{local,mgaberdiel}. Indeed,
\bea
Z(\tau,\bar\tau)
&=&
\chi_{{\cal V}_{-1/8}}\bar\chi_{{\cal V}_{-1/8}}
+\chi_{{\cal V}_{3/8}}\bar\chi_{{\cal V}_{3/8}}
+2\chi_{{\cal V}_0}\bar\chi_{{\cal V}_0}
+2\chi_{{\cal V}_0}\bar\chi_{{\cal V}_1}
+2\chi_{{\cal V}_1}\bar\chi_{{\cal V}_0}
+2\chi_{{\cal V}_1}\bar\chi_{{\cal V}_1}\nonumber\\
&=&
\frac{1}{\vert\eta(\tau)\vert^2}\sum_{k=0}^{3}\vert\Theta_{k,2}(\tau)\vert^2,
\eea
is easily verified to be invariant under both $\tau\rightarrow-1/\tau$ and
$\tau\rightarrow\tau+1$.


\subsection{Boundary states at $c=-2$}
As the Verlinde formula fails for the $c=-2$ triplet model, we can expect
a difficulty applying the method illustrated in Sec.2.3 to this model.
Indeed, when we try to find consistent boundary states as linear combinations
of Ishibashi states, we immediately come across ambiguity due to the degeneracy
of characters\cite{koganwheater,ishimoto}. 
On the other hand, since $c=-2$ theory is expected to model statistical systems
such as polymers, it is not conceivable that this theory has no consistent 
boundary states. 
In this subsection we see that such boundary states with consistent modular
properties can be found if we use the symplectic fermion representation of the 
triplet model\cite{kawaiwheater}.

Our starting point is to notice that the right hand side of the duality 
condition,
\beq
\sum_{i}n^i_{\tilde\alpha\tilde\beta}\chi_i(q)=
\langle\tilde\alpha\vert
(\tilde q^{1/2})^{L_0+\bar L_0-c/12}\vert\tilde\beta\rangle,
\label{eqn:cardycond}
\eeq
may be expanded with any basis of boundary states, not necessarily Ishibashi
states. 
Since Ishibashi states rely on well-defined conformal towers, 
we want to express consistent boundary states in terms of a more `sound' basis.
As the boundary must be diffeomorphism invariant,
the basis boundary states must satisfy
\beq
\left(L_m-\bar L_{-m}\right)|B\rangle =0.
\label{eqn:diffinv}
\eeq
In string theory, solutions to this condition are well known for boson, fermion
and ghost fields, and are found in the form of coherent states.
Let us find such coherent states for the symplectic fermions, following the
similar procedure for Majorana fermions\cite{callan}.
We shall demand vanishing of the boundary term in the action 
(\ref{eqn:action}). 
In terms of the mode operators of symplectic fermions, this leads to the
conditions on the boundary states,
\bea
&&\left(\chi^\pm_m-e^{\pm i\phi}\bar\chi^\pm_{-m}\right)\vert B\rangle=0,\\
&&\langle B\vert\left(\chi^\pm_m-e^{\pm i\phi}\bar\chi^\pm_{-m}\right)=0,
\eea
where $\phi$ is a phase factor reflecting the $U(1)$ symmetry of the system.
The states satisfying this condition are easily found to be
\bea
\vert B_{0\phi}\rangle 
&=& \exp
\left(
\sum_{k>0}\frac{e^{i\phi}}{k}\chi^-_{-k}\bar\chi^+_{-k}
         +\frac{e^{-i\phi}}{k}\bar\chi^-_{-k}\chi^+_{-k}
\right)
\vert 0_\phi\rangle,
\label{eqn:brabstate}\\
\langle B_{0\phi}\vert 
&=& \langle 0_\phi \vert \exp
\left(
\sum_{k>0}\frac{e^{i\phi}}{k}\chi^-_k\bar\chi^+_k
         +\frac{e^{-i\phi}}{k}\bar\chi^-_k\chi^+_k
\right),
\label{eqn:ketbstate}
\eea
where the vacua satisfy
\bea
&&\left(\chi^\pm_0-e^{\pm i\phi}\bar\chi^\pm_0\right)\vert 0_\phi\rangle=0,
\nonumber\\
&&\langle 0_\phi\vert\left(\chi^\pm_0-e^{\pm i\phi}\bar\chi^\pm_0\right)=0,
\label{eqn:0modecond}
\eea
and 
\bea
&&\chi_{n>0}^\pm\vert 0_\phi\rangle=0=\bar\chi_{n>0}^\pm\vert 0_\phi\rangle,\\
&&\langle 0_\phi\vert\chi_{n<0}^\pm=0=\langle 0_\phi\vert\bar\chi_{n<0}^\pm.
\eea
It can be verified that (\ref{eqn:brabstate}) and (\ref{eqn:ketbstate})
satisfy the condition (\ref{eqn:diffinv}) and its bra counterpart.

Just as the vanishing of energy-momentum flow across the boundary 
(\ref{eqn:t=tbar}) leads to the condition (\ref{eqn:diffinv}), 
one may consider the action of W-operators on the boundary,
\beq
\left[W^a-\Gamma\bar W^a\right]_{z=\bar z}=0,
\eeq
leading to the conditions
\beq
(W^a_m+\Gamma\bar W^a_{-m})|B\rangle =0,
\eeq
on the boundary states.
Here, $\Gamma$ is an element of a gluing automorphism group, which tells how 
the
holomorphic and antiholomorphic W-operators are related on the boundary.
Clearly, the simplest case is when the automorphism is trivial, $\Gamma=1$.
This restricts the value of the phase $\phi$ to be either $0$ or $\pi$.
We will see that this choice is sufficient to construct boundary states 
with consistent modular properties in the triplet model. 
When we deal with e.g. $Z_4$-orbifold symplectic fermion model, we need to 
consider
non-trivial automorphism as well as $\Gamma=1$.
The distinction of $\phi=0$ and $\pi$ corresponds to Dirichlet and Neumann
boundary conditions.
In the rest of this article we shall write $\phi=0$ as $+$ and $\phi=\pi$ as
$-$ for simplicity.

Now that we have explicit expressions of the boundary states, we may calculate 
the amplitudes between them, which appear on the right hand side of the 
consistency condition (\ref{eqn:cardycond}).

In the untwisted (or NS) sector of the triplet model, the vacua are doubly 
degenerate ($\omega$ and $\Omega$) and they may be normalised as
\bea
&&\langle\omega\vert\omega\rangle=\kappa,\\
&&\langle\Omega\vert\omega\rangle=-1,
\label{eqn:Omegaomega}\\
&&\langle\Omega\vert\Omega\rangle=0,
\eea
in accordance with the bulk theory\cite{sf}. 
Although the negative sign of (\ref{eqn:Omegaomega}) may look strange,
this is the choice of the sign which simplifies the following results
enormously.
As there are two vacua ($\omega$ and $\Omega$) each for $+$ and $-$ 
conditions, in the untwisted sector we have four boundary states
$\vert B_{\omega+}\rangle$, $\vert B_{\omega-}\rangle$, 
$\vert B_{\Omega+}\rangle$, and $\vert B_{\Omega-}\rangle$.
The amplitudes
$\langle a\vert (\tilde q^{1/2})^{L_0+\bar L_0+1/6}\vert b\rangle$ are
calculated using the mode operator algebra, as:\\

\begin{center}
\begin{tabular}{c|c|c|c|c}
\hline
&\multicolumn{4}{c}{$\vert b\rangle$}\\
\cline{2-5}
$\langle a\vert$&$B_{\omega+}$&$B_{\omega-}$&$B_{\Omega+}$&$B_{\Omega-}$\\
\hline\hline
$B_{\omega+}$&$(\kappa-\ln\tilde q)\eta(\tilde\tau)^2$&
$\;(\kappa-\ln\tilde q)\Lambda_{1,2}(\tilde\tau)\;$&
$-\eta(\tilde\tau)^2$&
$\;-\Lambda_{1,2}(\tilde\tau)\;$\\
\hline
$B_{\omega-}$&
$\;(\kappa-\ln\tilde q)\Lambda_{1,2}(\tilde\tau)\;$&
$(\kappa-\ln\tilde q)\eta(\tilde\tau)^2$&
$\;-\Lambda_{1,2}(\tilde\tau)\;$&
$-\eta(\tilde\tau)^2$\\
\hline
$B_{\Omega+}$&
$-\eta(\tilde\tau)^2$&
$-\Lambda_{1,2}(\tilde\tau)$&
$0$&$0$\\
\hline
$B_{\Omega-}$&
$-\Lambda_{1,2}(\tilde\tau)$&
$-\eta(\tilde\tau)^2$&$0$&$0$\\
\hline
\end{tabular}
\end{center}
We have denoted $\Lambda_{k,l}(\tau)=\Theta_{k,l}(\tau)/\eta(\tau)$.
An important point we would like to emphasise is that we cannot have all of 
the four states simultaneously. 
Because of the conditions (\ref{eqn:0modecond}), the amplitudes are
single-valued only when the bra and ket have the same value of $\phi$
(for example, $\langle 0_{\phi'}\vert\chi_0^\pm\vert 0_\phi\rangle
=e^{\pm i\phi}\langle 0_{\phi'}\vert\bar\chi_0^\pm\vert 0_\phi\rangle
=e^{\pm i\phi'}\langle 0_{\phi'}\vert\bar\chi_0^\pm\vert 0_\phi\rangle$, 
leading to $\phi=\phi'$).
Then in our triplet model, either $\vert B_{\omega+}\rangle$ 
or $\vert B_{\omega-}\rangle$ must be excluded.
Which one should be discarded is purely a matter of choice,
so for definiteness let us discard $\omega+$ in the following calculation.
The same thing happens when we construct the Ising model using Majorana
fermions, where a suitable GSO projection is necessary.

The amplitudes in the twisted (R) sector are rather straightforward.
The ground state is unique ($\mu$) and is normalised as
$\langle\mu\vert\mu\rangle=1$.
Corresponding to $+$ and $-$ conditions there are two boundary states
$\vert B_{\mu+}\rangle$ and $\vert B_{\mu-}\rangle$, and the amplitudes 
are:\\

\begin{center}
\begin{tabular}{c|c|c} 
\hline
&\multicolumn{2}{c}{$\vert b\rangle$}\\
\cline{2-3}
$\langle a\vert$&$B_{\mu+}$&$B_{\mu-}$\\
\hline\hline
$B_{\mu+}$&$\;\;\;\Lambda_{0,2}(\tilde\tau)-\Lambda_{2,2}(\tilde\tau)\;\;\;$&
$\;\;\;\Lambda_{0,2}(\tilde\tau)+\Lambda_{2,2}(\tilde\tau)\;\;\;$\\
\hline
$B_{\mu-}$&$\Lambda_{0,2}(\tilde\tau)+\Lambda_{2,2}(\tilde\tau)$&
$\Lambda_{0,2}(\tilde\tau)-\Lambda_{2,2}(\tilde\tau)$\\
\hline
\end{tabular}
\end{center}

We may now use these boundary-to-boundary amplitudes to expand the right hand 
side of the Cardy's condition
\beq
\sum_{i}n^i_{\tilde\alpha\tilde\beta}\chi_i(q)
=\sum_{a,b}\langle\tilde\alpha\vert a\rangle\langle a\vert
(\tilde q^{1/2})^{L_0+\bar L_0+1/6}
\vert b\rangle\langle b\vert\tilde\beta\rangle,
\label{eqn:cardyexp}
\eeq
with the coherent states.
Note that, as we are dealing with a non-unitary theory, 
$\langle\tilde\alpha\vert a\rangle$ and $\langle b\vert\tilde\beta\rangle$
must be treated merely as expansion coefficients (usual bra-ket operation does
not hold here).
Assuming the boundary state $\tilde{\cal V}_0$ associated to the vacuum 
representation satisfies 
$n^i_{\tilde\alpha\tilde{\cal V}_0}=n^i_{\tilde{\cal V}_0\tilde\alpha}
=\delta^i_{\tilde\alpha}$, we can follow the same procedure as Subsec.2.3. 
Comparing the coefficients on both sides of the equation, we find
\bea
&&\vert\tilde{\cal V}_0\rangle
=\frac{1}{2\sqrt{\pi}}\vert B_{\omega-}\rangle
+\frac{\kappa}{4\sqrt{\pi}}\vert B_{\Omega-}\rangle
+\frac 12 \vert B_{\mu-}\rangle,\nonumber\\
&&\vert\tilde{\cal V}_1\rangle
=\frac{-1}{2\sqrt{\pi}}\vert B_{\omega-}\rangle
-\frac{\kappa}{4\sqrt{\pi}}\vert B_{\Omega-}\rangle
+\frac 12 \vert B_{\mu-}\rangle,\nonumber\\
&&\vert\tilde{\cal V}_{-1/8}\rangle
=\vert B_{\mu+}\rangle
-2\sqrt{\pi}\vert B_{\Omega+}\rangle,\nonumber\\
&&\vert\tilde{\cal V}_{3/8}\rangle
=\vert B_{\mu+}\rangle
+2\sqrt{\pi}\vert B_{\Omega+}\rangle,\nonumber\\
&&\vert\tilde{\cal R}\rangle
\equiv\vert\tilde{\cal R}_0\rangle
=\vert\tilde{\cal R}_1\rangle
=2\vert B_{\mu-}\rangle.
\eea
These solutions are unique up to the $Z_2$ symmetry ($+\leftrightarrow -$),
and thus we can say that the consistent boundary states are unambiguously 
determined by the duality of open and closed string channels. 
If we substitute these boundary states back into the consistency equation
(\ref{eqn:cardyexp}), we can find possible multiplicity 
$n^i_{\tilde\alpha\tilde\beta}$ for each pair of boundary conditions.
It can be shown that $n^i_{\tilde\alpha\tilde\beta}$ coincide with the fusion
coefficients up to the ambiguity arising from the degeneracy of the character
functions.
 
What is non-trivial in this construction of the boundary states is the 
appearance of the term $\eta(\tilde\tau)^2\ln\tilde q$ in the untwisted sector 
amplitudes. 
Although the characters of the triplet model do not themselves close linearly
under the modular transformation, the zero-mode part
provides the $\eta(\tilde\tau)^2\ln\tilde q$ term which is needed for the
duality condition.
Note also that this term arises from the untwisted sector which is responsible 
for the logarithmic behaviour of the triplet model.
The difficulty we have experienced in finding Ishibashi states can be
understood as follows. After discarding the $\vert B_{\omega+}\rangle$ state,
the amplitudes in the untwisted sector become 
\[
\left(
\begin{array}{ccc}
(\kappa-\ln\tilde q)\eta(\tilde\tau)^2&-\Lambda_{1,2}(\tilde\tau)
&-\eta(\tilde\tau)^2\\
-\Lambda_{1,2}(\tilde\tau)&0&0\\
-\eta(\tilde\tau)^2&0&0
\end{array}
\right)
\]
which is obviously irregular.
One of the three eigenvalues of this matrix is zero, and the other two 
involve $\eta(\tilde\tau)^2\ln\tilde q$ which is not expressible using the
characters.
We therefore cannot find `Ishibashi' states which diagonalise the amplitudes
and give characters\footnote{Construction of generalised Ishibashi states is
discussed in \cite{bred1,bred2}.}.

We have seen that there exists a set of solutions to the Cardy's consistent
condition in the triplet model at $c=-2$, despite its pathological features
such as the failure of the Verlinde formula.
It is not at all clear whether these properties are shared by all or some of
the LCFT models, and it is not easy to analyse other models as this 
analysis relies on the particular free-field representation of the triplet 
model.
\newpage

\section{Conclusion}


In these lectures we have reviewed some basic ideas and methods of boundary 
CFT in the first part, and in the second part recent studies on its 
applications to logarithmic models have been discussed.
We have presented mainly two topics in the second part.
The first topic is on the behaviour of correlation functions in the presence of
boundary. We have seen in the $c=-2$ model the $2$-point functions of 
pre-logarithmic operators exhibit logarithmic divergence either far from or 
close to a boundary. 
The second topic is on the Cardy's consistency condition for logarithmic
CFTs. In the $c=-2$ triplet model, it is known that the character functions
do not transform linearly under the modular transformations, and what is 
more unusual, the Verlinde formula does not apply to this model.
Due to this pathological modular transformations of characters, we get into 
trouble if we attempt to express consistent boundary states as linear sums of 
Ishibashi states.
We have seen that nevertheless consistent boundary states do exist and may be 
expressed as linear combinations of coherent states constructed from free 
fields.

We would like to conclude this lecture note by mentioning a couple of open 
questions (actually LCFT is made of many open questions) and future prospects
which might be interesting.
The first is its relevance in applications to physical systems. 
There are indeed many examples of systems in statistical models and
string theory which have been claimed to be modelled by LCFTs. 
In regard of LCFT with boundary, besides `ordinary' and `extraordinary' 
transitions, the $O(n)$ model with $n<1$ has a `special' transition, which may
somehow be related to the boundary states of LCFT.
If a relation between some well-defined system and LCFT is established in
operator-content level, we can expect feed back from e.g. numerical simulations
and our understanding of LCFT would be accelerated enormously. 
The second point is the genericness of the claimed features of LCFTs.
So far the study of LCFT has been done on a case-by-case basis, 
and much of our 
understanding is based on specific free-field representations. 
If some sort of systematic classification (similar to ADE-type) is possible,
it may lead to more general discussions of LCFT.  
\newpage

\section*{Acknowledgements}
I would like to thank the Institute for Studies in Theoretical Physics and 
Mathematics 
(IPM), Tehran, Iran for their warm hospitality, and the organisers of the 
summer school, especially Shahin Rouhani, for hosting the productive and 
enjoyable meeting, and the secretaries of IPM for their efficient and kind 
arrangements.
I am grateful to Shahin Rouhani, Reza Rahimi Tabar, Matthias Gaberdiel, 
Michael Flohr, and many other participants of the school for fruitful 
discussions. 
I am also grateful to John Wheater for useful discussions and reading the 
manuscript, and Ian Kogan, John Cardy and Alexei Tsvelik for helpful 
conversations over the years.
I acknowledge University of Oxford for financial support.

\newpage
\appendix

\section{Ising model boundary states from Majorana fermions}
The free-field construction of the boundary states of the Ising model 
($c=1/2$) has been discussed by many authors\cite{nepo,yamaguchi}. 
As this is similar to our discussion for $c=-2$ theory (Sec.3.4) and
comparison may be helpful, we review the Ising model case in this appendix.

The system of the Majorana fermions with action
\beq
S=\frac{1}{2\pi}\int d^2 x(\bar\psi\del\bar\psi+\psi\bar\del\psi),
\label{eqn:isingaction}
\eeq
has central charge $1/2$ and realises the Ising model after Jordan-Wigner
transformation (which changes locality). 
The fermions are expanded as
\bea
&&\psi(\sigma, \tau)=\sum_n d_ne^{-n\tau+in\sigma},\\
&&\bar\psi(\sigma, \tau)=\sum_n \bar d_ne^{-n\tau-in\sigma},
\eea
and the vanishing of the boundary term in the action (\ref{eqn:isingaction})
implies conditions
\beq
(d_n\mp i\bar d_{-n}e^{2n\tau})\vert B\rangle=0.
\label{eqn:mbc}
\eeq
On the boundary at $\tau=0$ these conditions are satisfied by the states,
\beq
\vert B_\pm \rangle=\exp\left\{\sum_{n>0}\pm i d_{-n}\bar d_{-n}\right\}
\vert 0\rangle,
\eeq
where $\vert 0\rangle$ denotes the states annihilated by $d_{n>0}$ and
$\bar d_{n>0}$.
In the NS sector the vacuum is unique and we denote it as 
$\vert 0\rangle_{NS}$, whereas in the R sector we have doubly degenerate
vacua, $\vert \pm\rangle_{R}$.
These are treated symmetrically by demanding
\bea
&&d_0\vert\pm\rangle_R=\frac{1}{\sqrt{2}}e^{\pm i\pi/4}\vert\mp\rangle_R,\\
&&\bar d_0\vert\pm\rangle_R=\frac{1}{\sqrt{2}}e^{\mp i\pi/4}\vert\mp\rangle_R.
\eea
Thus we have four boundary states
\bea
&&\vert B_{\pm}\rangle_{NS}=\exp\left\{\pm i\sum_{n\geq 1/2}d_{-n}\bar d_{-n}
\right\}\vert 0\rangle_{NS},\nonumber\\
&&\vert B_{\pm}\rangle_{R}=\exp\left\{\pm i\sum_{n\geq 1}d_{-n}\bar d_{-n}
\right\}
\vert\pm\rangle_{R},
\label{eqn:ffbs}
\eea
where distinction of the two R ground states is absorbed.
We may now calculate cylinder (boundary-to-boundary) amplitudes
(in the order of $\vert B_+\rangle$, $\vert B_-\rangle$) as\\
$\langle a\vert (\tilde q^{1/2})^{L_0+\bar L_0-c/12}\vert b\rangle=$
\beq
\mbox{NS:}
\left(
\begin{array}{cc}
\sqrt{\frac{\theta_3(\tilde\tau)}{\eta(\tilde\tau)}}&
\sqrt{\frac{\theta_4(\tilde\tau)}{\eta(\tilde\tau)}}\\
\sqrt{\frac{\theta_4(\tilde\tau)}{\eta(\tilde\tau)}}&
\sqrt{\frac{\theta_3(\tilde\tau)}{\eta(\tilde\tau)}}
\end{array}
\right),
\eeq
\beq\mbox{R:}
\left(
\begin{array}{cc}
\sqrt{\frac{\theta_2(\tilde\tau)}{2\eta(\tilde\tau)}}&
0\\
0&
\sqrt{\frac{\theta_2(\tilde\tau)}{2\eta(\tilde\tau)}}
\end{array}
\right).
\eeq
We cannot have the both states of R sector in order to make the theory 
consistent. 
Discarding $\vert B_-\rangle_R$, we may use
$\vert B_+\rangle_{NS}$, $\vert B_-\rangle_{NS}$, $\vert B_+\rangle_R$
as the basis of the boundary states. 

The Ising model has identity ($I$), energy ($\epsilon$) and spin ($\sigma$) 
density operators.
The characters for the three representations are
\bea
&&\chi_I=\lishi I \vert 
(\tilde q^{1/2})^{L_0+\bar L_0-c/12}\vert I\rishi
=\frac 12 \sqrt{\frac{\theta_3(\tilde\tau)}{\eta(\tilde\tau)}}
+\frac 12 \sqrt{\frac{\theta_4(\tilde\tau)}{\eta(\tilde\tau)}},
\label{eqn:charI}\\
&&\chi_\epsilon=\lishi\epsilon \vert 
(\tilde q^{1/2})^{L_0+\bar L_0-c/12}\vert \epsilon\rishi
=\frac 12 \sqrt{\frac{\theta_3(\tilde\tau)}{\eta(\tilde\tau)}}
-\frac 12 \sqrt{\frac{\theta_4(\tilde\tau)}{\eta(\tilde\tau)}},
\label{eqn:chare}\\
&&\chi_\sigma=\lishi \sigma \vert 
(\tilde q^{1/2})^{L_0+\bar L_0-c/12}\vert\sigma\rishi
=\frac 12 \sqrt{\frac{\theta_2(\tilde\tau)}{\eta(\tilde\tau)}},
\label{eqn:chars}
\eea
where 
$\vert I\rishi$, $\vert\epsilon\rishi$, $\vert\sigma\rishi$
are the Ishibashi states for $I$, $\epsilon$, $\sigma$, respectively.

We may compare both sides of the consistency condition,
assuming the existence of the state $\vert\tilde 0\rangle$ such that 
$n^i_{\tilde 0\tilde\alpha}=n^i_{\tilde\alpha\tilde 0}
=\delta^i_{\tilde\alpha}$,
and find consistent boundary states expressed as linear sum of the basis 
states. 
If we expand with the Ishibashi states, we have the result of Subsec.2.3,
\bea
&&\vert\uparrow\rangle=\vert\tilde I\rangle
=2^{-1/2}\vert I\rishi+2^{-1/2}\vert\epsilon \rishi
+2^{-1/4}\vert\sigma \rishi,\\
&&\vert\downarrow\rangle=\vert\tilde\epsilon\rangle
=2^{-1/2}\vert I\rishi+2^{-1/2}\vert\epsilon \rishi
-2^{-1/4}\vert\sigma \rishi,\\
&&\vert F\rangle=\vert\tilde\sigma\rangle
=\vert I\rishi-\vert\epsilon \rishi.
\eea
Which is up and which is down in the first two lines is a matter of choice. 
This procedure is equally carried out using the coherent states of Majorana
fermions, and we find,
\bea
&&\vert\uparrow\rangle
=2^{-1/2}\vert B_+\rangle_{NS}+2^{-1/4}\vert B_+\rangle_R,\\
&&\vert\downarrow\rangle
=2^{-1/2}\vert B_+\rangle_{NS}-2^{-1/4}\vert B_+\rangle_R,\\
&&\vert F\rangle
=\vert B_-\rangle_{NS}.
\eea
The Ising model has two $Z_2$ symmetries,
namely, high-temperature-low-temperature symmetry (P) and
spin up-down symmetry (Q).
Actions of P and Q on the boundary states are
\bea
P:&& \frac{\vert\uparrow\rangle+\vert\downarrow\rangle}{\sqrt{2}}
\leftrightarrow\vert F\rangle,\\
Q:&& \vert\uparrow\rangle\leftrightarrow\vert\downarrow\rangle,\;
\vert F\rangle\leftrightarrow\vert F\rangle.
\eea
In terms of the Majorana fermions, P exchanges 
$\vert B_+\rangle$ and $\vert B_-\rangle$.
Q is blind about $\pm$, but flips the sign of $\vert B\rangle_R$.
\newpage

\section{Elliptic modular functions}

We list some formulae of elliptic modular functions.
The Jacobi theta functions used in Sec. 3.4 are defined as
\beq
\Theta_{\lambda,\mu}(\tau)=\sum_{k\in Z}q^{(2\mu k+\lambda)^2/4\mu},
\eeq
and those used in App. A are
\bea
&&\theta_2(\tau)=\sum_{k\in Z}q^{(k+1/2)^2/2},\\
&&\theta_3(\tau)=\sum_{k\in Z}q^{k^2/2},\\
&&\theta_4(\tau)=\sum_{k\in Z}(-1)^kq^{k^2/2},
\eea
where $q=e^{2\pi i\tau}$.
As is easily verified,
\bea
\theta_2(\tau)&=&2\Theta_{1,2}(\tau),\\
\theta_3(\tau)&=&\Theta_{0,2}(\tau)+\Theta_{2,2}(\tau),\\
\theta_4(\tau)&=&\Theta_{0,2}(\tau)-\Theta_{2,2}(\tau).
\eea
The Dedekind eta function is defined as
\beq
\eta(\tau)=q^{1/24}\prod_{n=1}^\infty(1-q^n).
\eeq
In Sec. 3.4 we have used the notation
\beq
\Lambda_{\lambda,\mu}(\tau)=\frac{\Theta_{\lambda,\mu}(\tau)}{\eta(\tau)}.
\eeq
Under the modular T ($\tau\rightarrow\tau+1$) and S ($\tau\rightarrow-1/\tau$)
transformations, $\Theta_{\lambda,\mu}(\tau)$ and $\eta(\tau)$ transform as
\bea
\Theta_{\lambda,\mu}(-1/\tau)
&=&\sqrt{\frac{-i\tau}{2\mu}}\sum_{\nu=0}^{2\mu-1}e^{\lambda\nu\pi i/\mu}
\Theta_{\nu,\mu}(\tau),\nonumber\\
\eta(-1/\tau)&=&\sqrt{-i\tau}\eta(\tau),
\label{eqn:modularS}
\eea
and
\bea
\Theta_{\lambda,\mu}(\tau+1)
&=&e^{\lambda^2\pi i/2\mu}\Theta_{\lambda,\mu}(\tau),\nonumber\\
\eta(\tau+1)&=&e^{\pi i/12}\eta(\tau).
\label{eqn:modularT}
\eea
The three functions $\theta_2(\tau)$, $\theta_3(\tau)$ and $\theta_4(\tau)$
transform each other under T and S, as
\bea
\theta_2(\tau+1)=\sqrt i\theta_2(\tau),\\
\theta_3(\tau+1)=\theta_4(\tau),\\
\theta_4(\tau+1)=\theta_3(\tau),
\eea
and
\bea
\theta_2(-1/\tau)=\sqrt{-i\tau}\theta_4(\tau),\\
\theta_3(-1/\tau)=\sqrt{-i\tau}\theta_2(\tau),\\
\theta_4(-1/\tau)=\sqrt{-i\tau}\theta_3(\tau).
\eea
The modular S matrix for the Ising model used in Sec.2.3 is, 
in the order of $I$, $\epsilon$, $\sigma$,
\beq
S_{ij}=\frac 12\left(
\begin{array}{ccc}
1&1&\sqrt 2\\
1&1&-\sqrt 2\\
\sqrt 2&-\sqrt 2&0
\end{array}
\right),
\eeq
which may be verified from the characters of Ising model
(\ref{eqn:charI}) - (\ref{eqn:chars}) using the above transformation formulas.
\newpage


\end{document}